\documentstyle[12pt]{article}

\begin{document}

\titlepage

{\it University of Shizuoka}

\hspace*{9.5cm} {\bf US-95-07}\\[-.3in]

\hspace*{9.5cm} {\bf August 1995}\\[.3in]


\begin{center}

{\Large\bf  Democracy of Families }\\[.2in]

{\Large\bf  and }\\[.2in]

{\Large\bf  Neutrino Mass Matrix }\footnote{ 
Talk persented at the YITP Workshop on {\it Flavor Physics and 
Beyond the Standard Model}, YITP, Kyoto University, Japan,  July 26 - 28, 
1995.  The work was, in part, reported at the IV International 
Symposium on {\it Weak and Electromagnetic Interactions in Nuclei}, 
Osaka, Japan, June 12-16, 1995}\\[.5in]

{\bf Yoshio Koide}\footnote{
E-mail: koide@u-shizuoka-ken.ac.jp} \\

Department of Physics, University of Shizuoka \\ 
395 Yada, Shizuoka 422, Japan \\[.5in]

{\large\bf Abstract}\\[.1in]

\end{center}

\begin{quotation}
On the basis of a seesaw-type mass matrix model for quarks and leptons, 
$M_f \simeq m_L M_F^{-1} m_R$, where $m_L\propto m_R$ are universal 
for $f=u,d,\nu$ and $e$ (up-quark-, down-quark-, neutrino- and charged 
lepton-sectors), and $M_F$ is given by $M_F=K ({\bf 1} + 3 b_f X)$ 
({\bf 1} is a $3\times 3$ unit matrix, $X$ is a democratic-type matrix 
and $b_f$ is a complex parameter which depends on $f$, neutrino mass 
spectrum and mixings 
are discussed.  The model can provide an explanation why $m_t \gg m_b$, 
while $m_u\sim m_d$ by taking $b_u=-1/3$, at which the detarminant of 
$M_F$ becomes zero.  At $b_\nu=-1/2$, the model can provide a large 
$\nu_\mu$-$\nu_\tau$ mixing, $\sin^2 2\theta_{23}\simeq 1$, with 
$m_{\nu 1} \ll m_{\nu 2} \simeq m_{\nu 3}$, which is favorable to the 
atmospheric and solar neutrino data.
\end{quotation}

\newpage
\begin{center}
{\Large\bf Democracy of Families \\[.1in]

and Neutrino Mass Matrix}
\end{center}

\hfill {\bf Yoshio Koide} \ \ \ 
{\bf  (University of Shizuoka)} 

\vglue.2in

{\large\bf 1. Basic standpoint}


Considering the rapid increasing of the mass spectrum of quarks 
and leptons as seen in Fig.~1, usually the horizontal degree of 
freedom is called ``generations". 
The term ``generations" suggests that there are hierarchical 
differences among those generations.
However, here, I would like to use a term ``families" for the 
horizontal degree of freedom. 

\begin{picture}(420,200)

\vspace{8cm}
\end{picture}

\begin{center}
{\bf Fig.1 \ \ Quark and lepton mass spectrum.}
\end{center}

My standpoint is as follows: 
All families take equivalent positions among them, i.e., 
there is no family with a special position in the original 
state.
This does not always mean that 
the families should, for example, described by SU(3) symmetry.

For example, if it is possible,
I would like to describe 
these mass matrices only in terms of 
``unit matrix and democratic-type matrix [1], i.e.,
$$
{\bf 1}=\left(
\begin{array}{ccc}
1 & 0 & 0 \\
0 & 1 & 0 \\
0 & 0 & 1 
\end{array} \right) \ \ \  {\rm and}\  \ \  X=\frac{1}{3}\left(
\begin{array}{ccc}
1 & 1 & 1 \\
1 & 1 & 1 \\
1 & 1 & 1 
\end{array} \right) \ .\eqno(1.1)
$$
Unfortunately, the real mass spectrum (Fig.~1) seems to be 
more complicated to describe in terms of (1.1) only.

\vglue.2in

{\large\bf 2.\ A unified quark and lepton mass matrix model}


Before discussing my neutrino mass matrix, I would like to give 
a short review of a unified quark and lepton mass matrix model.

We consider 
vector-like heavy fermions $F_i$ 
in addition to quarks and leptons $f_i$
$(f_i=u_i, d_i, e_i,\nu_i ; \ i=1,2,3)$.
These fermions belong to $F_L=(1,1)$, $F_R=(1,1)$, 
$f_L=(2,1)$, and $f_R=(1,2)$ of SU(2)$_L \times$SU(2)$_R$, 
respectively.
We assume the mass matrix form[2] for $(f, F)$ 
$$
M=\left(
\begin{array}{ccc}
0 & m_L \\
m_R & M_F
\end{array} \right) \ ,  \eqno(2.1)
$$
where we assume that the chiral symmetry breaking terms 
$m_L\propto m_R$ have a universal structure for quarks and leptons
and the heavy fermion mass term 
$M_F$ has a structure of (unit matrix)+(a rank-one matrix) and   
$M_F$ includes only one complex parameter 
which depends on quarks or leptons, and up- or down-,
as we state later.
 As well-known, the $6\times 6$ mass matrix (2.1) leads to the so-called 
seesaw form 
$$
M_f\simeq m_L M_F^{-1} m_R \ , \eqno(2.2)
$$
for ${\rm Tr}M_F \gg {\rm Tr}m_R,  {\rm Tr}m_L$.

For the origin of the mass matrices $m_L$ and $m_R$, 
I would like to consider a U(3)-family nonet Higgs potential 
scenario, which leads to an excellent charged lepton mass relation [3]
$$
m_e+m_\mu+m_\tau=\frac{2}{3}(\sqrt{m_e}+\sqrt{m_\mu}+\sqrt{m_\tau})^2
\ , \eqno(2.3)
$$
However, such a multi-Higgs model, in general, induces 
flavor-changing neutral currents. 
The phenomenological study of the constraints on the 
Higgs boson masses has been given in Ref.~[4] in the 
collaboration with Tanimoto:
we have  estimated that $m_H \sim $ a several TeV  
from $\Delta m(K_S$-$K_L)$ , $\cdots$, 
and rare decays $K_L\rightarrow e^\pm + \mu^\mp$, $\cdots$.

However, since I have no sufficient time to review the 
scenario, I would like to skip the review from the 
present talk.
Hereafter, apart from this scenario, we assume simply 
$$
m_L \propto m_R \propto M_e^{1/2}\equiv 
{\rm diag}(\sqrt{m_e}, \sqrt{m_\mu}, \sqrt{m_\tau}) \ .
\eqno(2.4)
$$

The most exciting feature of the present quark mass matrix 
is as follows:
the model can naturally understand that 
why $m_t\gg m_b$, while $m_u \sim m_d$,  
without introducing such a parameter as it takes 
a large value in up-quark sector compared with 
that in down-quark sector.

The basic idea is as follows: we assume the following 
form [5]  of $M_F$,
$$
M_F\propto O_F= {\bf 1} + 3 b_f e^{i\beta_f} X  \ ,
\eqno(2.5)
$$
then the inverse matrix of $O_F$ is given by
$$
O_F^{-1}={\bf 1} + 3 a_f e^{i\alpha_f}X \ , \eqno(2.6)
$$
with 
$$
a_f e^{i\alpha_f} =- \frac{b_f e^{i\beta_f}}{1+3b_f e^{i\beta_f}} \ .
\eqno(2.7) 
$$

Why $m_t \gg m_b$ can be understood by taking $b_u=-1/3$ ($\beta_u=0$) 
because  $b_u\rightarrow -1/3$ provides $|a_u|\rightarrow \infty$,
so that top-quark mass enhancement is caused (note that the seesaw 
form (2.2) is not valid any longer in the limit of  
$b_u\rightarrow -1/3$).
Why $m_u\sim m_d$ is understood from the fact that 
democratic mass matrix makes only the third family heavy, 
i.e., the effects of $|a_u|\rightarrow\infty$ affects only to $m_t$.

\begin{picture}(420,250)

\end{picture}

\begin{quotation}
{\bf Fig.2 \ \ Mass spectrum versus a parameter $b_f$:}
solid and broken lines denote the cases of $\beta_f = 0$ and 
$\beta_f = -20^{\circ}$, respectively.
\end{quotation}

The behavior of the mass spectrum versus $b_fe^{i\beta_f}$ is 
given in Fig.~2, where parameters $k$ and $K_f$ are defined by
$$
M=\left(
\begin{array}{ccc}
0 & m_L \\
m_R & M_F
\end{array} \right) =
m_0 \left(
\begin{array}{cc}
0 &  Z \\
k Z & K_f O_F \\
\end{array}
\right) \ ,   \eqno(2.8)
$$
$$
Z=\left(
\begin{array}{ccc}
z_1 & 0 & 0 \\
0 & z_2 & 0 \\
0 & 0 & z_3 
\end{array} \right)  \ , \ \ 
O_F=\left(
\begin{array}{ccc}
1 & 0 & 0 \\
0 & 1 & 0 \\
0 & 0 & 1 
\end{array} \right) + b_f e^{i\beta_f}
\left(
\begin{array}{ccc}

1 & 1 & 1 \\
1 & 1 & 1 \\
1 & 1 & 1 
\end{array} \right) \ . \eqno(2.9)
$$

Note that from the phenomenological point of view, it is not essential 
that $O_F=$[(unit matrix)+(democratic-type matrix)].
Instead of (2.9), we may take
$$
O_F=\left(
\begin{array}{ccc}
1 & 0 & 0 \\
0 & 1 & 0 \\
0 & 0 & 1 
\end{array} \right) + 3 b_f e^{i\beta_f}
\left(
\begin{array}{ccc}

0 & 0 & 0 \\
0 & 0 & 0 \\
0 & 0 & 1 
\end{array} \right) \ . 
\eqno(2.10)
$$
However, then, we must take the matrix $Z$ as the 
non-diagonal form
$$
Z=\frac{1}{6} \left( 
\begin{array}{ccc}
3(z_2+z_1) &  -\sqrt{3}(z_2-z_1) & -\sqrt{6}(z_2-z_1) \\
 -\sqrt{3}(z_2-z_1) & 4z_3+z_2+z_1 & -\sqrt{2}(2z_3 -z_2-z_1) \\
-\sqrt{6}(z_2-z_1) &  -\sqrt{2}(2z_3 -z_2-z_1) & 2(z_3+z_2+z_1) \\
\end{array} \right) \ . \eqno(2.11) 
$$
One may consider that the matrix forms (2.10) and (2.11) are 
favorable to model-building.
However, I believe that the forms (2.9) are more promising.

Taking ($b_u = -1/3$, $\beta_u=0$) and 
($b_d=-1.0$, $\beta_d=-18^\circ$) , but keeping $K_u=K_d$, 
we can provide not only reasonable quark mass ratios 
$m_u/m_c$, $m_c/m_t$, $m_d/m_s$ and $m_s/m_b$, but also 
$m_u/m_d$, $m_c/m_s$ and $m_t/m_b$, and, moreover, 
we can provide reasonable values of Kobayashi-Maskawa (KM) matrix 
parameters.
For the details, please see a preprint Ref.[6] 
in the collaboration with Fusaoka.

\newpage

{\large\bf 3.  Neutrino mass matrix  
with large $\nu_\mu$-$\nu_\tau$ mixing} 

So far, we have assumed the following mass terms:
$$ 
\nu_L=(2,1,3)_L^{Y=-1} 
\hspace*{3.3cm}
\nu_R=(1,2,3)_R^{Y=-1}
$$
$$
\hspace*{-1.5cm}\updownarrow \ m_L
\hspace{5.5cm}\updownarrow \ m_R
$$
$$
N_R=(1,1,3)_R^{Y=0} 
\hspace*{1cm} \longleftarrow\longrightarrow \hspace*{1cm}
N_L=(1,1,3)_L^{Y=0} \ \ .
$$
$$
\hspace*{0.5cm}
M_F \ \  \eqno(3.1)
$$
Now, in order to understand why $m_\nu \ll m_\ell, m_q$,
we must introduce a large Majorana mass term $M_M$ 
($\gg M_D$) (hereafter, we denote the Dirac mass term 
$M_F$ for $F=N$ as $M_D$ in contrast to the Majorana 
mass matrix $M_M$).

\begin{center} Table I. Comparison between Model I and Model II 
\end{center}
\begin{tabular}{cc}\hline
\hspace*{2cm}{\rm Model I}\hspace*{2cm} 
& \hspace*{3cm}{\rm Model II}\hspace*{3cm} \\ \hline
   &    \\
{\rm $L$-$R$ Symmetric} & {\rm $L$-$R$ Asymmetric} \\
   &    \\
$N_L$ {\rm and} $N_R$ {\rm acquire} $M_M$ & $\nu_R$ {\rm acquires} $M_M$ \\
\end{tabular}
$$
\begin{array}{cc}
M = \left(
\begin{array}{cccc}
0 & 0 & 0 & \frac{1}{2}m_L \\
0 & 0 & \frac{1}{2}m_R^T & 0 \\
0 & \frac{1}{2}m_R & M_M & M_D \\
\frac{1}{2}m_L^T & 0 & M_D^T & M_M 
\end{array}
 \right)
 & 
M = \left(
\begin{array}{cccc}
0 & 0 & 0 & \frac{1}{2}m_L \\
0 & M_M & \frac{1}{2}m_R^T & 0 \\
0 & \frac{1}{2}m_R & 0 & M_D \\
\frac{1}{2}m_L & 0 & M_D^T & 0 
\end{array} \right) \\
   &    \\
M_{\nu_L} \simeq \left(\frac{1}{2}\right)^2 m_L M_M^{-1} m_L^T  & 
\begin{array}{c}
M_{\nu_L} \simeq \left(\frac{1}{2} \right)^4 m_L M_D^{-1} m_R M_M^{-1} \\
\times m_R^T (M_D^T)^{-1} m_L^T
\end{array} \\
   &    \\
{\rm Assume}\ M_M = \frac{K_M}{K_D}M_D & 
{\rm Assume}\ M_M = m_0 K_M {\bf 1} \\
   &    \\
M_{\nu_L} \simeq \frac{1}{4} \frac{m_0}{K_M} Z O_F^{-1}Z & 
M_{\nu_L} \simeq \frac{1}{16} \frac{k^2 m_0}{K_D^2 K_M}
Z O_F^{-1} Z \cdot Z O_F^{-1}Z  \\ \hline
\end{array}
$$ 

I would like to propose two models:
Model I, in which the vector-like heavy fermions $N_L$ and $N_R$ 
acquire large Majorana masses $M_M$, respectively; 
Model II, in which the right-handed neutrinos $\nu_R$ acquire 
large Majorana masses $M_M$.
In the model I, the chiral SU(2)$_R$ symmetry is broken by $m_R$, 
while, in the model II, it is broken by $M_M$ with an extremely 
large energy scale. 
The characteristics of the models are listed in Table I. 
The predicted values of the neutrino mixings are identical in the 
models I and II, although the predicted values of neutrino mass 
ratios are different from each other.
Note that in the model I, the light ``right-handed" Majorana 
neutrinos $\nu'_{Ri}$, which are originated from $\nu_{Ri}$, 
appear with masses $m(\nu'_{Ri})\simeq k^2 m(\nu_{Ri})$ 
($i=1,2,3$) (we suppose $k \geq 10$).

\vspace{.1in}

We show the typical cases of $b_\nu$ (for simplicity, we 
consider the case $\beta_\nu=0$) for the model I.

\noindent
[{\bf Case $b_\nu=-\frac{1}{3}+\varepsilon$}]

$$ 
m_{\nu 1}\simeq \frac{3}{8} \frac{m_e}{m_\tau} \frac{m_0}{K_M} \ , \ \ \ 
m_{\nu 2}\simeq \frac{1}{2}\frac{ m_\mu}{m_\tau} \frac{m_0}{K_M} \ , \ \ \ 
m_{\nu 3}\simeq \frac{1}{27\sqrt{2}}\frac{m_0}{|\varepsilon| K_M}
 \ ,  \eqno(3.3)
$$
where $m_0=\sqrt{3} m_t$ at $\mu=\Lambda_W=175$ GeV.
$$
U_{\nu L} \simeq \left(
\begin{array}{ccc}
1 & -\frac{1}{2}\sqrt{m_e/m_\mu} & -\frac{1}{2}\sqrt{m_e/m_\tau} \\
\frac{1}{2}\sqrt{m_e/m_\mu} & 1 & -\sqrt{m_\mu/m_\tau} \\
\sqrt{m_e/m_\tau} & \sqrt{m_\mu/m_\tau} & 1 
\end{array} \right) \ .\eqno(3.4)
$$

\noindent
[{\bf  Case $b_\nu = -1/2$}]

$$ 
m_{\nu 1}\simeq \frac{1}{2} \frac{m_e}{m_\tau} \frac{m_0}{K_M} \ , \ \ \ 
m_{\nu 2}\simeq m_{\nu 3} \simeq 
\frac{1}{4} \sqrt{\frac{m_\mu}{ m_\tau}} \frac{m_0}{K_M} \ ,  \eqno(3.5)
$$
$$
U_{\nu L} \simeq \left(
\begin{array}{ccc}
1 & -\sqrt{m_e/m_\mu} & -\sqrt{m_e/m_\tau} \\
\frac{1}{\sqrt{2}}\left(\sqrt{\frac{m_e}{m_\mu}}- 
\sqrt{\frac{m_e}{m_\tau}}\right) & \frac{1}{\sqrt{2}} & 
-\frac{1}{\sqrt{2}} \\
\frac{1}{\sqrt{2}}\left(\sqrt{\frac{m_e}{m_\mu}}+ 
\sqrt{\frac{m_e}{m_\tau}}\right) & \frac{1}{\sqrt{2}} & 
\frac{1}{\sqrt{2}} \\
\end{array} \right) \ .\eqno(3.6)
$$

\noindent
[{\bf Case $b_\nu = -1$}]

$$ 
m_{\nu 1}\simeq m_{\nu 2}\simeq  
\frac{1}{4} \sqrt{\frac{m_e m_\mu}{m_\tau^2}} \frac{m_0}{K_M} \ , \ \ \ 
m_{\nu 3}= \frac{1}{8}  \frac{m_0}{K_M} \ ,  \eqno(3.7)
$$
$$
U_{\nu L} \simeq \left(
\begin{array}{ccc}
\frac{1}{\sqrt{2}} & \frac{1}{\sqrt{2}} & 
\frac{1}{\sqrt{2}} \left( \sqrt{\frac{m_\mu}{m_\tau}}
-\sqrt{\frac{m_e}{m_\tau}}\right)  \\
-\frac{1}{\sqrt{2}} & \frac{1}{\sqrt{2}} & 
\frac{1}{\sqrt{2}} \left( \sqrt{\frac{m_\mu}{m_\tau}}
+\sqrt{\frac{m_e}{m_\tau}}\right)  \\
-\sqrt{m_e/m_\tau} & -\sqrt{m_\mu/m_\tau} & 1 
\end{array} \right) \ .\eqno(3.8)
$$

The $b_\nu$-dependency of the mass spectrum of the light neutrinos, 
($m_{\nu1}, m_{\nu2}, m_{\nu3}$), is similar to that in Fig.2. 
The $b_\nu$-dependency of the neutrino mixing matrix $U_{ij}$ is 
given in Fig.3.

\begin{picture}(420,270)

\end{picture}

\begin{center}
{\bf Fig.3 \ \ $U_{ij}$ versus $b_\nu$.}
\end{center}

Recent atmospheric neutrino data from the Kamiokande [7] have suggested 
that the $\nu_\mu$-$\nu_X$ mixing ($X = e$ or $\tau$) 
is caused maximally, i.e., $\sin^2 2\theta \simeq 1$, with 
$\Delta m^2 \sim 10^{-2}$ eV. On the other hand, 
solar neutrino data [8] have suggested that 
$\sin^2 2 \theta \simeq 7 \times 10^{-3}$ and 
$\Delta m^2 \simeq 6 \times 10^{-6} {\rm eV^2}$. 
We consider that the atmospheric neutrino data show 
$\nu_\mu$-$\nu_\tau$ mixing, while the solar neutrino data show 
$\nu_e$-$\nu_\mu$ mixing, so that we interests in the case of 
$b_\nu \simeq -1/2$, which provides $\sin^2 2 \theta_{23} \sim 1$ 
with $m_{\nu1} \ll m_{\nu2} \simeq m_{\nu3}$.

Although, by taking $b_\nu \simeq -1/2$, we can get 
$\sin^2 2\theta_{e \mu} \simeq 7 \times 10^{-3}$ and 
$\sin^2 2 \theta_{\mu \tau} \sim 1$, 
in the model I, we cannot explain the experimental fact 
$\Delta m^2_{32}/\Delta m^2_{21} \sim 10^3$, 
because $m_{\nu2}$ and $m_{\nu3}$ are highly degenerated at 
$b_\nu \simeq -1/2$ (at most, $\Delta m^2_{32}/\Delta m^2_{21} \sim 10^2$ 
at $b_\nu \simeq -0.4$). The situation cannot improved even if 
we consider $\beta_\nu \neq 0$.

In the model II, if we take, for example, $b_\nu = -0.41$, 
we can get the predictions $\sin^2 2\theta_{12} = 6.8 \times 10^{-3}$, 
$\Delta m^2_{21} \equiv 6 \times 10^{-6} {\rm eV^2}$, 
$\sin^2 2\theta_{23} = 0.58$, 
$\Delta m^2_{32} = 0.99 \times 10^{-2} {\rm eV^2}$, 
$m_{\nu1} = 2.4 \times 10^{-8} {\rm eV}$, 
$m_{\nu2} = 2.5 \times 10^{-3}$ eV, 
$m_{\nu3} = 0.10$ eV, where $\Delta m^2_{21} = 6 \times 10^{-6} {\rm eV^2}$ 
has been taken as an input value in order to fix the  free parameter $K_M$.
Similarly, the case of $b_\nu = -0.5$ with $\beta_\nu \simeq 10^{\circ}$ 
can give a simultaneous explanation of the atmospheric and solar 
neutrino data.

Also note that the case $b_\nu = -0.5$ with $\beta_\nu = 0$ can 
give a rough explanation of the atmospheric and LSND neutrino data 
(the LSND data [9] have suggested that 
$\sin^2 2\theta_{12} \simeq 3 \times 10^{-3}$ and 
$\Delta m^2_{21} \simeq 6 \ {\rm eV^2}$), 
because the case predicts that 
$\sin^2 2\theta_{12} = 7.7 \times 10^{-3}$, 
$\Delta m^2_{21} = 1.7 \ {\rm eV^2}$, 
$\sin^2 2 \theta_{23} = 0.99 \times 10^{-2}$, 
$\Delta m^2_{32} \equiv 1.6 \times 10^{-2}$ eV, 
$m_{\nu1} = 1.7 \times 10^{-4}$ eV, $m_{\nu2} = 1.3$ eV, 
$m_{\nu3} = 1.3$ eV. If this picture is correct, the mass matrix in 
the lepton sector must rigorously real, because 
the case $\beta_\nu \neq 0$ 
makes the mass degeneration between $\nu_2$ and $\nu_3$ mild.

\vglue.2in

{\large\bf 4. Summary}

As an extension of a unified quark and lepton mass matrix 
(2.8) with (2.9),
neutrino masses and their mixings have investigated.
When we take $b_e=0$, reasonable values of up- and down-quark mass
ratios and KM matrix parameters are obtained 
by taking $(b_u=-1/3,\ \beta_u=0)$ and 
$(b_d=-1,\ \beta_d\simeq 18^\circ)$, and 
with keeping $K_u=K_d$.

For neutrino mass matrix, we have proposed two models:
In Model I, $N_L$ and $N_R$ acquire large Majorana 
masses $M_M$ ($\propto M_N$), so that SU(3)$_{family}$ 
is badly broken by the energy scale of $M_M$; 
In Model II, $\nu_R$ acquire  large Majorana masses 
$M_M\propto {\bf 1}$, so that SU(2)$_R$ is badly broken 
by the energy scale of $M_M$.
In the both models, especially, the case $b_\nu\simeq -1/2$ is 
interesting, because the case  provides 
a maximal mixing $\sin^2 2\theta_{23}\simeq 1$ 
together with $m_{\nu 2}\simeq m_{\nu 3}$.
Phenomenologically, Model II is favorable to the atmospheric 
and solar neutrino data.

In the present stage, why nature choose $b_\nu \simeq -1/2$, 
$b_e=0$, $b_u=-1/3$ and $b_d\simeq -1$ is an open question.
The phenomenological success of the present unified mass matrix 
form (2.8) with (2.9) should be taken seriously, and 
a more plausible model-building must be investigated urgently.

\vspace{.3in}



\newcounter{0000}
\centerline{\large\bf References }
\begin{list}
{[~\arabic{0000}~]}{\usecounter{0000}
\labelwidth=0.8cm\labelsep=.1cm\setlength{\leftmargin=0.7cm}
{\rightmargin=.2cm}}
\item H.~Harari, H.~Haut and J.~Weyers, Phys.~Lett. {\bf B78},
459 (1978);
T.~Goldman, in {\it Gauge Theories, Massive Neutrinos and 
Proton Decays}, edited by A.~Perlumutter (Plenum Press, New York, 
1981), p.111;
T.~Goldman and G.~J.~Stephenson,~Jr., Phys.~Rev. {\bf D24}, 
236 (1981); 
Y.~Koide, Phys.~Rev.~Lett. {\bf 47}, 1241 (1981); Phys.~Rev. 
{\bf D28}, 252 (1983); {\bf 39}, 1391 (1989);
C.~Jarlskog, in {\it Proceedings of the International Symposium on 
Production and Decays of Heavy Hadrons}, Heidelberg, Germany, 1986
edited by K.~R.~Schubert and R. Waldi (DESY, Hamburg), p.331, 1986;
P.~Kaus, S.~Meshkov, Mod.~Phys.~Lett. {\bf A3}, 1251 (1988); 
Phys.~Rev. {\bf D42}, 1863 (1990);
L.~Lavoura, Phys.~Lett. {\bf B228}, 245 (1989); 
M.~Tanimoto, Phys.~Rev. {\bf D41}, 1586 (1990);
H.~Fritzsch and J.~Plankl, Phys.~Lett. {\bf B237}, 451 (1990); 
Y.~Nambu, in {\it Proceedings of the International Workshop on 
Electroweak Symmetry Breaking}, Hiroshima, Japan, (World 
Scientific, SIngapore, 1992), p.1.
\item The seesaw mechanism has originally proposed for the purpose of 
explaining why neutrino masses are so invisibly small:
M.~Gell-Mann, P.~Rammond and R.~Slansky, in {\it Supergravity}, 
edited by P.~van Nieuwenhuizen and D.~Z.~Freedman (North-Holland, 
1979); T.~Yanagida, in {\it Proc. Workshop of the Unified Theory and 
Baryon Number in the Universe}, edited by A.~Sawada and A.~Sugamoto 
(KEK, 1979); R.~Mohapatra and G.~Senjanovic, Phys.~Rev.~Lett. 
{\bf 44} (1980)  912.

For  applications of the seesaw mechanism 
to the quark mass matrix, see, for example, 
Z.~G.~Berezhiani, Phys.~Lett. {\bf 129B} (1983)  99;
Phys.~Lett. {\bf 150B} (1985)  177;
D.~Chang and R.~N.~Mohapatra, Phys.~Rev.~Lett. {\bf 58} 
1600 (1987); 
A.~Davidson and K.~C.~Wali, Phys.~Rev.~Lett. {\bf 59} (1987)  393;
S.~Rajpoot, Mod.~Phys.~Lett. {\bf A2} (1987)  307; 
Phys.~Lett. {\bf 191B} (1987)  122; Phys.~Rev. {\bf D36} (1987)  1479;
K.~B.~Babu and R.~N.~Mohapatra, Phys.~Rev.~Lett. {\bf 62}  (1989) 1079; 
Phys.~Rev. {\bf D41} (1990)  1286;  
S.~Ranfone, Phys.~Rev. {\bf D42} (1990)  3819; 
A.~Davidson, S.~Ranfone and K.~C.~Wali, Phys.~Rev. 
{\bf D41} (1990)  208; 
I.~Sogami and T.~Shinohara, Prog.~Theor.~Phys. {\bf 66}  (1991)  1031;
Phys.~Rev. {\bf D47}  (1993)  2905; 
Z.~G.~Berezhiani and R.~Rattazzi, Phys.~Lett. {\bf B279}  (1992)  124;
P.~Cho, Phys.~Rev. {\bf D48} (1994)  5331; 
A.~Davidson, L.~Michel, M.~L,~Sage and  K.~C.~Wali, Phys.~Rev. 
{\bf D49} (1994)  1378; 
W.~A.~Ponce, A.~Zepeda and R.~G.~Lozano, Phys.~Rev. {\bf D49} 
 4954 (1994).
\item Y.~Koide, Phys.~Rev. {\bf D49} (1994)  2638.
\item Y.~Koide and M.~Tanimoto, US-95-04 and EHU-95-03, (1995) 
(hep-ph/9505333).
\item A seesaw neutrino mass matrix with a democratic-ype matrix 
$M_F$ was first discussed by Goldman and Stephenson, although  
their interest was not in quark mass matrix, but in neutrino 
mass matrix, and their matrix $M_F$ did not include the unit 
matrix part, differently from the form (2.5): 
T.~Goldman and G.~J.~Stephenson,~Jr., Phys.~Rev. {\bf D24}, 
236 (1981).
\item  Y.~Koide and H.~Fusaoka, US-95-03 and AMU-95-04 (1995) 
(hep-ph/9505201).
\item Y.~Fukuda {\it et al}, Phys.~Lett. {\bf B335}, 237 (1994).
\item GALLEX collaboration, P.~Anselmann {\it et al}, 
Phys.~Lett. {\bf B327}, 377 (1994);

SAGE collaboration, J.~N.~Abdurashitov {\it et al}, 
Phys.~Lett. {\bf B328}, 234 (1994);
\item C.~Athanassupoulos {\it et al}, LA-UR-95-1238 (1995) 
(nucl-ex/9504002).
\end{list}

\end{document}